\documentclass[11pt]{article}
\usepackage{amsfonts,amsthm,amsmath,amssymb,upgreek}
\usepackage[paper=letterpaper,margin=1in]{geometry}
\usepackage[export]{adjustbox}
\usepackage[numbers,sort&compress]{natbib}
\usepackage{color}
\usepackage{tikz}
\usepackage{graphicx}% Include figure files
\usepackage{dcolumn}% Align table columns on decimal point
\usepackage{bm}% bold math
\usepackage{hyperref}% add hypertext capabilities
\usepackage{float}
\usepackage[caption=false]{subfig}
\usepackage{hyperref}
\hypersetup{
    colorlinks=true,
    linktoc=page,    %set to all if you want both sections and subsections linked
    linkcolor=blue,
    urlcolor=magenta
}

\newcommand{\be}{\begin{equation}}
\newcommand{\ee}{\end{equation}}
\newcommand{\bea}{\begin{eqnarray}}
\newcommand{\eea}{\end{eqnarray}}
\newcommand{\dd}{\text{d}}
\newcommand{\nn}{\nonumber}
\newcommand{\bra}[1]{\left<{#1}\right|}
\newcommand{\ket}[1]{\left|{#1}\right>}
\newcommand{\ave}[1]{\left<{#1}\right>}
\newcommand{\tr}{\text{tr}}% representing trace
\newcommand{\tN}{\text{N}}

\newcommand{\hh}{\text{h}}
\newcommand{\hH}{\text{H}}
\newcommand{\tE}{\text{E}}
\newcommand{\eff}{\text{eff}}

\newcommand{\daggerfootnote}[1]{%
    \renewcommand{\thefootnote}{\fnsymbol{footnote}}%
    \footnote[2]{#1}
    \renewcommand{\thefootnote}{\arabic{footnote}}%
}

\begin{document}
%===============================

\thispagestyle{empty}

\vspace*{.5cm}
\begin{center}

{\bf {\Large Quantum-corrected black hole thermodynamics\\
from the gravitational path integral
}\\
\vspace{1cm}}

\begin{center}

 {\bf Yu-Qi Liu$^{a}$, Hao-Wei Yu$^{a,b}$, and Peng Cheng$^{a}$\daggerfootnote{p.cheng.nl@outlook.com}}\\
  \bigskip \rm
  
\bigskip
 a) Center for Joint Quantum Studies and Department of Physics, \\School of Science, Tianjin University, Tianjin 300350, China\\
b) School of Physical Science and Technology, \\ Southwest University, Chongqing 400715, China

\rm
  \end{center}

\vspace{1.5cm}
{\bf Abstract}
\end{center}
\begin{quotation}
Exploring quantum effects from black hole thermodynamics has always been a pivotal topic. In recent years, the free energy landscape and ensemble-averaged theory based on the Euclidean path integral approach have provided further understanding of the statistical aspects of the black hole system. We investigate the quantum-corrected thermodynamics of the Reissner-Nordstr\"om AdS black hole by including off-shell geometries in a reduced gravitational path integral. Within this collective-variable approximation, we derive an effective action by considering the subleading-order terms in the ensemble-averaged theory and show that the corresponding thermodynamic quantities define a consistent thermodynamics. Furthermore, the phase diagram was modified by the off-shell effects, resulting in a more abundant phase structure. We show that the traditional black hole thermodynamics can be recovered in the semi-classical limit. The region of first-order phase transitions shrinks and zero-order phase transitions emerge when off-shell effects are included. These results provide a tractable framework for understanding how off-shell black hole geometries generate quantum corrections to black hole phase structures.
\end{quotation}

\vspace{1cm}

\setcounter{page}{0}
\setcounter{tocdepth}{2}
\setcounter{footnote}{0}

\newpage
{\noindent} \rule[-10pt]{16.5cm}{0.05em}\\
\tableofcontents
{\noindent} \rule[-10pt]{16.5cm}{0.05em}\\
%\pagebreak

%%%%%%%%%%%%%%%%%%%%%%%%%%%%%%%%%%%%%%%%%%%%%%%%%%%%%%%%%%%%%%%%%%%%%%%%%%%%%%%%%%%%%%%%%%%%%%%%%%%%
% MAIN BODY
%%%%%%%%%%%%%%%%%%%%%%%%%%%%%%%%%%%%%%%%%%%%%%%%%%%%%%%%%%%%%%%%%%%%%%%%%%%%%%%%%%%%%%%%%%%%%%%%%%%%

%\tableofcontents

%%%%%%%%%%%%%%%%%%%%%%%%%%%%%%%%%%%%%%%%%%%%%%%%%%%%%%%%%%%%%%%%%%%%%%%%%%%%%%%%%%%%%%%%%%%%%%%%%%%%

\section{Introduction}
\label{intro}

%%%%%%%%%%%%%%%%%%%%%%%%%%%%%%%%%%%%%%%%%%%%%%%%%%%%%%%%%%%%%%%%%%%%%%%%%%%%%%%%%%%%%%%%%%%%%%%%%%%%

Black hole thermodynamics has always been an intriguing subject since its birth in the 1970s~\cite{Bekenstein:1973ur,Bardeen:1973gs,Bekenstein:1975tw}, which has largely improved our understanding of black hole microstructure~\cite{Hawking:1982dh,Strominger:1996sh,Maldacena:1996gb,Wei:2015iwa} and holographic principle~\cite{tHooft:1993dmi,Susskind:1994vu,Witten:1998zw,Gross:1980he,Sundborg:1999ue}.
In asymptotically Anti-de Sitter (AdS) spacetime, black holes were discovered to have a rich tapestry of interesting phase structures.
The Hawking-Page transition between thermal AdS and Schwarzschild black hole~\cite{Hawking:1982dh} corresponds to the confinement/deconfinement phase transition in the dual large $N$ SU($N$) gauge theory on a sphere~\cite{Witten:1998qj,Witten:1998zw}.
For the Reissner-Nordstr\"om (RN)-AdS black holes, the phase transition between large and small black holes resembles the phase behavior of a Van der Waals fluid, with the same critical exponents~\cite{Chamblin:1999hg,Chamblin:1999tk,Kastor:2009wy,Dolan:2011xt,Kubiznak:2012wp}.

As one of the most exciting developments in recent years, the free energy landscape proposal \cite{Li:2020khm,Li:2020nsy,Li:2020spm,Yang:2021ljn,Li:2021vdp,Li:2022oup,Li:2022yti,Li:2022ylz} provided a unique angle to understand black hole phase transitions based on the generalized free energy.
The generalized free energy can be derived by including conical singularities in the Euclidean gravitational path integral ~\cite{Gibbons:1976ue,York:1986it,Braden:1987ad,Whiting:1988qr,York:1988va}.
When interpreting the Euclidean path integral as calculating the statistical aspect of the system and ensemble-averaging over all the possible states,% with the density matrix derived from the path integral, 
a further statistical understanding of the black hole thermodynamics was obtained \cite{Cheng:2024hxh,Cheng:2024efw,Ali:2024adt}. 
%Recently, off-shell thermodynamic methods have also been further developed in the context of higher-curvature black holes, where the horizon radius is used to parametrize thermodynamic branches and the off-shell Gibbs free energy provides a natural tool for diagnosing equilibrium and local stability~\cite{Hao:2026cco}.

It is natural to expect that when off-shell geometries with conical singularity are included in the path integral, the ensemble-averaged quantities are corrected by the off-shell effects. The corrected quantities should recover the black hole thermodynamics result in the semi-classical limit.
Moreover, the ensemble-averaged quantities should define a valid thermodynamics, and we hope they can provide more insights into the phase structure.
In this paper, our main focus is on the ensemble-averaged thermodynamics. 
The questions we want to answer are whether quantum-corrected thermodynamics gives rise to a well-defined thermodynamic description, and how the off-shell geometries affect the phase structure of the RN-AdS black hole.

By studying the ensemble-averaged thermodynamics in more detail, we have gained further understanding of black hole thermodynamics.
Evaluating the ensemble-averaged theory up to the subleading-order correction of $G_\tN$, we get an effective action, from which all the thermodynamic quantities can be obtained.
The physical quantities derived from the ensemble-averaged theory should be regarded as quantum-corrected quantities, and we show that those quantities define a valid thermodynamic system.
The phase diagram of the RN-AdS black hole is corrected by the off-shell effects, and we have a richer phase structure, including a zero-order phase transition in the corrected thermodynamics.
More specifically, in the case of a first-order phase transition between small and large black holes, this transition tends to happen at a lower ensemble temperature for larger dimensionless Newton's constant $G_\tN$.
We use the typical size of the system to define the dimensionless $\tilde{G}_\tN$ and drop the tilde sign for convenience \cite{Cheng:2024hxh}; thus, different values of $G_\tN$ actually mean different ratios between the typical size and the Planck length. 

%\textcolor{red}{Evaluating the ensemble-averaged theory up to the subleading order in the reduced horizon-radius sector, we obtain an effective action for the collective variable $r_\hh$, from which the corresponding thermodynamic quantities can be derived. The physical quantities obtained in this way should be regarded as quantum-corrected effective quantities associated with the off-shell black hole geometries included in the reduced path integral. We show that these quantities define a consistent effective thermodynamic system.}

The paper is organized as follows. In Sec. \ref{review}, we systematically review the idea of using ensemble-averaged theory to understand black hole thermodynamics. We exhibit the expression of the density matrix and the probability distribution of different states, and show the semi-classical limit of the ensemble averaged theory and the quantum correction.
Sec. \ref{thermdy} describes the effective thermodynamics, where we derive all the physical quantities from the ensemble-averaged theory and check the first law of thermodynamics.
Sec. \ref{sec:PT} is dedicated to demonstrating the phase structure in the quantum-corrected thermodynamics. We show how the off-shell geometries contribute to the phase transition for finite $G_\tN$.
%We discuss more general situations in Sec. \ref{general}.
We conclude the paper and make further discussions in Sec. \ref{con}.

%%%%%%%%%%%%%%%%%%%%%%%%%%%%%%%%%%%%%%%%%%%%%%%%%%%%%%%%%%%%%%%%%%%%%%%%%%%%%%%%%%%%%%%%%%%%%%%%%%%%

\section{The ensemble-averaged theory}
\label{review}

%%%%%%%%%%%%%%%%%%%%%%%%%%%%%%%%%%%%%%%%%%%%%%%%%%%%%%%%%%%%%%%%%%%%%%%%%%%%%%%%%%%%%%%%%%%%%%%%%%%%

In this section, we review recent developments on deriving black hole thermodynamics from an ensemble-averaged theory inherited from the gravitational Euclidean path integral \cite{Cheng:2024hxh,Cheng:2024efw,Ali:2024adt}.
By including more Euclidean geometries with conical singularity in the phase space, we can calculate the density matrix and probability distribution of the geometric configurations via the Euclidean path integral.
This means we are including more off-shell geometries in the gravitational path integral, and it turns out that the semi-classical approximation of the theory naturally gives rise to the black hole thermodynamics of the RN-AdS black hole.

\subsection{Density matrix and probability distribution}

\begin{figure}[hbt]
\centering
  \includegraphics[width=6cm]{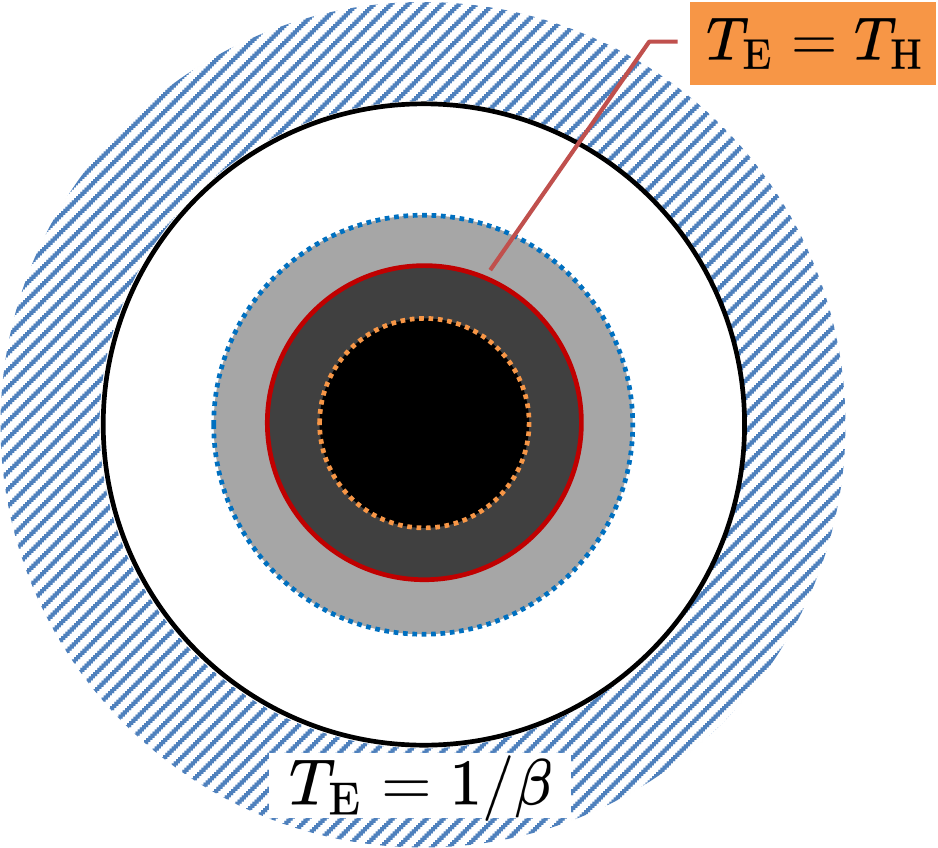}
  \caption{Gravitational configurations with a given ensemble temperature $T_{\text{E}}$.}
  \label{ensembleT}
\end{figure}

Now, let us consider the following physical setup, as illustrated in Fig. \ref{ensembleT}.
Given an asymptotically AdS cavity with a fixed ensemble temperature $T_{\text{E}}$, if the whole system is in equilibrium, what bulk geometries can be filled for a given bulk gravity?
The semi-classical results are well-known.
For bulk Einstein-Hilbert gravity, the answer is the famous Hawking-Page transition \cite{Hawking:1982dh}; when charges are included, there can be phase transitions between large and small black holes \cite{Chamblin:1999hg,Chamblin:1999tk,Kastor:2009wy,Dolan:2011xt,Kubiznak:2012wp}. 

The semi-classical results are far from enough.
We have only considered fixed on-shell bulk configurations, which are too monotonous and dull. 
It is natural to include more off-shell geometries in the phase space, when one wants to explore a little bit of the quantum gravitational effect of the black hole system.
In this paper, we mainly focus on the system with given ensemble temperature $T_\text{E}$, charge $Q$, and pressure $P\propto \Lambda$.
Inspired by the black hole free energy landscape proposal \cite{Li:2020khm,Li:2020nsy,Li:2020spm,Yang:2021ljn,Li:2021vdp,Li:2022oup,Li:2022yti,Li:2022ylz}, the most important off-shell ingredients that should be included in the phase space are geometries of different sizes, shown in Fig. \ref{ensembleT}. 
The use of $r_\hh$ as the label of the off-shell configurations is a collective-variable approximation. This approximation is motivated by the fact that the small/large black hole phase transition is primarily characterized by the black hole size. The horizon radius distinguishes the relevant thermodynamic branches and therefore serves as a natural order parameter for the phase structure studied below. We do not claim that this reduced description exhausts the full gravitational path integral. 
Rather, the free energy landscape captures the dominant saddle structure of the gravitational path integral by connecting the relevant on-shell and nearby off-shell black hole configurations through the horizon radius collective coordinate.
Recent off-shell thermodynamic studies further support using the horizon radius as a natural parameter for black-hole branches and stability analysis~\cite{Hao:2026cco}.
Alternatively, one can interpret it as a black hole with a quantum width \cite{Marolf:2003bb,Marolf:2018ldl}.
Then the question left is how to evaluate the weight of those gravitational configurations in the phase space.

The Euclidean path integral method provides a natural way to evaluate the density matrix and a statistical interpretation of thermodynamics.
For scalar fields, the density matrix and partition function can be derived through the Euclidean path integral, imitating the scattering amplitude formula, we have
\bea
\bra{\phi_b}e^{-\beta H}\ket{\phi_a}=\int_{\phi(\textbf{x},0)=\phi_a(\textbf{x})}^{\phi(\textbf{x},\beta)=\phi_b(\textbf{x})}[\mathcal{D} \phi]~ e^{-I_{\text{E}}}\,,\label{density}
\eea
and
\bea
Z_{\phi}=\sum_{\phi}\bra{\phi}e^{-\beta H}\ket{\phi}=\int_{\text{periodic}}[\mathcal{D} \phi]~ e^{-I_{\text{E}}}\,,\label{Zphi}
\eea
with Euclidean action $I_\text{E}$ and periodic $\tau$ direction.
The representation implies that the weight of each state can be evaluated through the path integral, as far as the Euclidean action can be evaluated.

The same logic can be generalized to gravity theory \cite{Gibbons:1976ue,York:1986it}, and we can obtain the density matrix through the Euclidean gravitational path integral.
The off-shell geometries have a conical singularity on the tip of their Euclidean counterpart, which is a reflection of their off-shell property.
For RN-AdS black holes, the Euclidean action can be evaluated as \cite{Li:2022oup}
\be
I_{\tE}[\begin{matrix}
\begin{tikzpicture}
\draw[thick] (0.5/2,0) .. controls (0.5/2,0) and (0.25/2,0.025/2) .. (0,0.25/2);
\draw[thick] (0.5/2,0.5/2) .. controls (0.5/2,0.5/2) and (0.25/2,1.9/8) .. (0,0.25/2);
\draw[thick] (0.5/2,0.25/2) ellipse (0.125/2 and 0.25/2);
\fill[red] (0,0.25/2) circle (1pt);
\end{tikzpicture}
\end{matrix}]=\frac{\beta r_\hh}{2G_\tN}\left(1+\frac{r_\hh^2}{L^2}+\frac{Q^2}{r_\hh^2}-2\pi r_\hh T_\tE \right)\,,
\ee
with AdS radius $L$ and inverse temperture $\beta=1/T_\tE$.
Then, the density matrix can be represented as \cite{Cheng:2024hxh}
\be
\rho=\sum_{
\begin{matrix}
\begin{tikzpicture}
\draw[thick] (-0.05,0) .. controls (-0.05,0) and (-0.175,0.0125) .. (-0.3,0.125);
\draw[thick] (-0.05,0.25) .. controls (-0.05,0.25) and (-0.175,1.9/8) .. (-0.3,0.125);
\draw[thick] (-0.05,0.125) ellipse (0.125/2 and 0.25/2);
\fill[red] (-0.3,0.125) circle (1pt);
\end{tikzpicture}
\end{matrix}
}
e^{-I_\tE[
\begin{matrix}
\begin{tikzpicture}
\draw[thick] (0.5/2,0) .. controls (0.5/2,0) and (0.25/2,0.025/2) .. (0,0.25/2);
\draw[thick] (0.5/2,0.5/2) .. controls (0.5/2,0.5/2) and (0.25/2,1.9/8) .. (0,0.25/2);
\draw[thick] (0.5/2,0.25/2) ellipse (0.125/2 and 0.25/2);
\fill[red] (0,0.25/2) circle (1pt);
\end{tikzpicture}
\end{matrix}
]}
\ket{
\begin{matrix}
\begin{tikzpicture}
\draw[thick] (0.5,0) .. controls (0.5,0) and (0.25,0.025) .. (0,0.25);
\draw[thick] (0.5,0.5) .. controls (0.5,0.5) and (0.25,1.9/4) .. (0,0.25);
\draw[thick] (0.5,0.25) ellipse (0.125 and 0.25);
\fill[red] (0,0.25) circle (1pt);
\end{tikzpicture}
\end{matrix}
}
\bra{
\begin{matrix}
\begin{tikzpicture}
\draw[thick] (-0.5,0) .. controls (-0.5,0) and (-0.25,-0.025) .. (0,-0.25);
\draw[thick] (-0.5,-0.5) .. controls (-0.5,-0.5) and (-0.25,-1.9/4) .. (0,-0.25);
\draw[thick] (-0.5,-0.25) ellipse (-0.125 and -0.25);
\fill[red] (0,-0.25) circle (1pt);
\end{tikzpicture}	
\end{matrix}
},
\ee
and we have the partition function being defined as
\be
Z=\tr ~\rho\,.
\ee
With the density matrix and partition function, the normalized probability distribution of each state in the phase space can be written as
\be
\hat{P}[\begin{matrix}
\begin{tikzpicture}
\draw[thick] (0.5/2,0) .. controls (0.5/2,0) and (0.25/2,0.025/2) .. (0,0.25/2);
\draw[thick] (0.5/2,0.5/2) .. controls (0.5/2,0.5/2) and (0.25/2,1.9/8) .. (0,0.25/2);
\draw[thick] (0.5/2,0.25/2) ellipse (0.125/2 and 0.25/2);
\fill[red] (0,0.25/2) circle (1pt);
\end{tikzpicture}
\end{matrix}]
=\frac{e^{-I_\tE[\begin{matrix}
\begin{tikzpicture}
\draw[thick] (0.5/2,0) .. controls (0.5/2,0) and (0.25/2,0.025/2) .. (0,0.25/2);
\draw[thick] (0.5/2,0.5/2) .. controls (0.5/2,0.5/2) and (0.25/2,1.9/8) .. (0,0.25/2);
\draw[thick] (0.5/2,0.25/2) ellipse (0.125/2 and 0.25/2);
\fill[red] (0,0.25/2) circle (1pt);
\end{tikzpicture}
\end{matrix}]}}{Z}\,.
\ee

The shape of the probability distribution depends on the value of $G_\tN$. 
Restricting ourselves to 4-dimensional cases, one can always use the typical size of the system to redefine a dimensionless Newton constant $\tilde{G}_\tN$ and drop the tilde sign for convenience. 
The $G_\tN$ dependence of the distribution is illustrated in Fig. \ref{Pdistribution}. 
All physical results should be ensemble-averaged, obtained by averaging over all states in the phase space with the probability distribution shown in Fig. \ref{Pdistribution}, i.e.
\be
\bar{A}=\ave{A}=\tr (\rho A)\,.\label{eq:ave}
\ee

\subsection{The semi-classical limit and beyond}
\label{IIsemi}

\begin{figure*}[hbt]
\centering
%\subfloat[Gibbs free energy]{\includegraphics[width=0.75\linewidth]{Gibbs.pdf}}\\
\subfloat[Probability distribution with $G_\tN=1/1000$ \label{Pd-a}]{\includegraphics[width=0.45\linewidth]{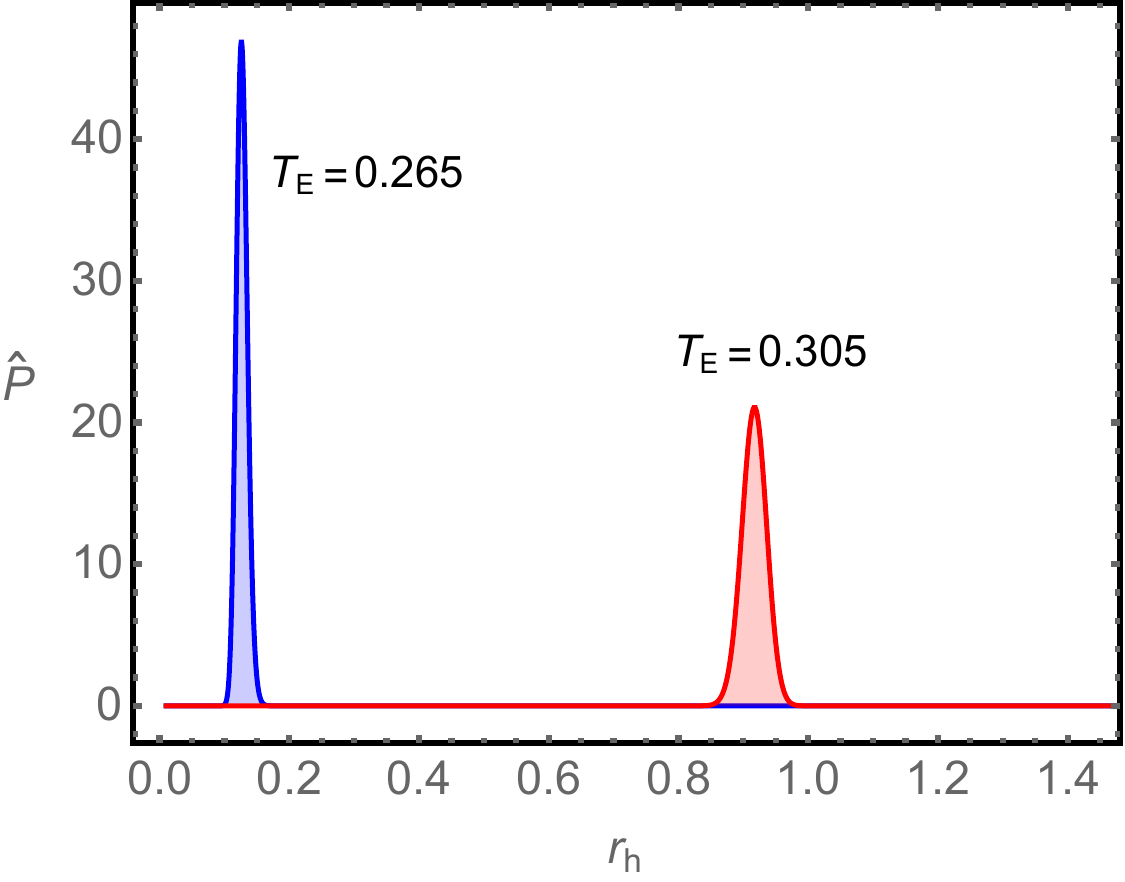}}~~~~~~~~
\subfloat[Probability distribution with $G_\tN=1/10$\label{Pd-b}]{\includegraphics[width=0.45\linewidth]{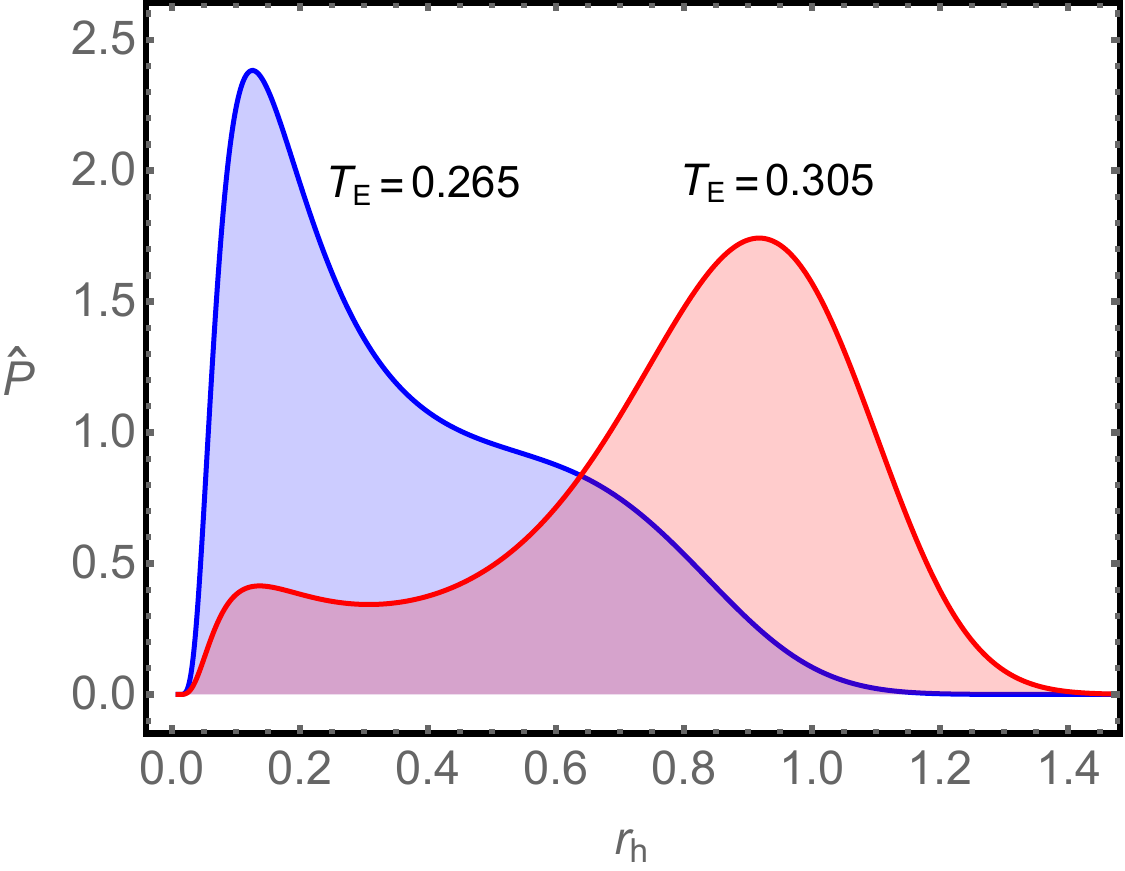}}
    \caption{Probability distributions for the RN-AdS spacetime ($Q=0.1$ and $L=1$) with $G_\tN=1/1000$ and $G_\tN=1/10$. The blue curves represent the small black hole phase with $T_\tE=0.265$; While for the red curves, we have the large black hole phase with $T_\tE=0.305$.}
    \label{Pdistribution}
\end{figure*}

Let us look closely at the probability distribution shown in Fig. \ref{Pdistribution}.
As can be seen from the figures, for a relatively small value of Newton's constant $G_\tN=1/1000$, the probability distribution is dominated by the on-shell configurations and can be approximated by a Gaussian distribution with means at $r_\hH=0.126$ and $r_\hH=0.917$, respectively. 
For extremely small $G_\tN \to 0$, the distribution would be a Dirac delta function, which only picks up a single geometry when doing average or path integral.
All the semi-classical black hole physics with a fixed bulk configuration can be derived in the limit.
For instance, the sharp black hole phase transitions can be directly derived as $G_\tN \to 0$ \cite{Cheng:2024hxh}. This should also be regarded as the thermodynamic limit of the statistical theory.
For relatively large $G_\tN=1/10$, the off-shell geometries away from the classical result are making significant contributions, as can be seen from Fig. \ref{Pd-b}. 
Consequently, the ensemble-averaged result deviates significantly from the classical result.

For finite $G_\tN$, there are always corrections from the off-shell geometries.
For the case shown in Fig. \ref{Pd-b}, one can numerically integrate over the probability distribution. Moreover, for the Gaussian distribution shown in Fig. \ref{Pd-a}, we can analytically integrate over the distribution and obtain the leading-order correction from the geometries not far from the classical saddle.
This correction is most study-worthy when the size of the system is larger than the Planck length, yet the Planckian physical effects should not be ignored.

%%%%%%%%%%%%%%%%%%%%%%%%%%%%%%%%%%%%%%%%%%%%%%%%%%%%%%%%%%%%%%%%%%%%%%%%%%%%%%%%%%%%%%%%%%%%%%%%%%%%

\section{Quantum-corrected black hole thermodynamics from off-shell geometries}
\label{thermdy}

%%%%%%%%%%%%%%%%%%%%%%%%%%%%%%%%%%%%%%%%%%%%%%%%%%%%%%%%%%%%%%%%%%%%%%%%%%%%%%%%%%%%%%%%%%%%%%%%%%%%

For the case when the system size is larger than the Planck length but the Planckian physical effects can not be ignored, we can use a Gaussian distribution to approximate the probability distribution shown in Fig. \ref{Pd-a}.
In this section, we employ the probability distribution to clarify the implications of off-shell geometries for black hole thermodynamics.
%In this section, we are going to use the distribution to understand what the off-shell geometries mean for black hole thermodynamics.

First of all, let us consider the ensemble with given temperature $T_\tE$, charge $Q$, and pressure $P$ more carefully. 
The partition function should be written as an integral over the energy of each state $E$ as
\be
Z=\int e^{-\beta E} D(E)\dd E\,,\label{eq:8}
\ee
with density of state $D(E)$. 
Note that $D(E)$ denotes the number of quantum states per unit energy interval with energy $E$, which is the exponential of the Bekenstein-Hawking entropy.
It can be further written as 
\be 
Z=\int e^{-\beta E+S(E)} \dd E%=\int e^{-\beta F} \dd E
=\int e^{-I_\tE (E)} \dd E\,.
\ee
Thus, in Eq. \eqref{eq:ave}, the parameter that is used to label different states and thus needs to be averaged over should be the energy (mass) of the spacetime $E$. $E(r_\hh)$ denotes the energy of the spacetime with horizon radius $r_\hh$. Although a conical singularity in their Euclidean counterpart presents, the bare energy $E(r_\hh)$ can be expressed as
\be
E(r_\hh)=\frac{r_\hh}{2G_\tN}\left(1+\frac{r_\hh^2}{L^2}+\frac{Q^2}{r_\hh^2}\right)\,.
\ee
As shown in Eq. \eqref{eq:ave}, all the physical quantities should be ensemble averaged with respect to the probability distribution shown in Fig. \ref{Pdistribution}.
For relatively small $G_\tN$, using the Gaussian distribution to approximate the distribution, means we can expand everything to the second order of $\delta=r_\hh-r_\hH$.
$r_\hH$ is the on-shell horizon radius, and for the case shown in Fig. \ref{Pd-a} we have $r_\hH=0.126$ and $r_\hH=0.917$.
The most direct quantity to be averaged over is the energy, and we have
\bea
\bar{E}=\ave{E} &=& \frac{1}{Z}\int e^{-I_\tE} (\partial_\beta I_\tE)~( \partial_{r_\hh} E)~\dd r_\hh \nn\\
&\approx & \frac{\partial I_\tE}{\partial \beta}+\frac{1}{2 I_{\tE}''}\cdot \frac{\partial I_{\tE}''}{\partial \beta} + \mathcal{O}(G_\tN^1)\nn\\
&=& \partial_\beta \left[ I_\tE+\frac{1}{2} \ln I_{\tE}'' \right]+ \mathcal{O}(G_\tN^1)\,,\label{barE-1}
\eea
where the prime represents derivative with respect to $r_\hh$.
In the second line, we have used the Gaussian distribution to approximate the probability distribution. Dropping the subsubleading-order terms, we can further rewrite the averaged energy as
\bea
\bar{E}=\frac{r_\hH}{2G_\tN}\left(1+\frac{r_\hH^2}{L^2}+\frac{Q^2}{r_\hH^2}\right) +\frac{L^2 Q^2+3 r_\hH^4}{2L^2 \left(\beta  Q^2-2 \pi  r_\hH^3\right)+6 \beta  r_\hH^4}%+ \frac{L^4 Q^2 \left(r_\hH^2-Q^2\right)+3 L^2 r_\hH^6+9 r_\hH^8}{4 \pi  L^2 r_\hH^3 \left(L^2 \left(3 Q^2-r_\hH^2\right)+3 r_\hH^4\right)}
\label{barE-2}
\eea
It can be easily seen from Eq. \eqref{barE-2}, the leading order contribution that is proportional to $1/G_\tN$ is exactly the semi-classical result of the black hole thermodynamics.
The subleading correction proportional to $G_\tN ^0$ may play an important role for relatively large $G_\tN$.
From Eq. \eqref{barE-1}, we get an effective action
\be
\Gamma_{\eff}=I_\tE+\frac{1}{2} \ln I_{\tE}''\,,
\ee
and the partition function can be approximated as
\be
Z\approx e^{-\Gamma_{\eff}}\,,
\ee
in the subleading-order approximation.

Note that the tree-level contribution is determined by the classical saddle, while subleading terms in the $G_\tN$ expansion capture loop effects in the path integral description. The finite-$G_\tN$ effects considered here can therefore be regarded as loop corrections to black hole thermodynamics within the reduced phase space. We analytically study the one-loop correction to black hole thermodynamics in this restricted phase space; higher-order corrections in the reduced path integral deserve further investigation. 
In this sense, the subleading finite-$G_\tN$ contribution should be viewed as a quantum correction associated with the collective fluctuation of off-shell black hole geometries.
%Note that the tree-level scattering amplitude is proportional to vertices, and subleading-order terms of $G_\tN$ capture the loop corrections in the possible quantum theory of gravity.
%So the finite $G_\tN$ effects in the Euclidean path integral that we are trying to capture can be regarded as loop corrections to the black hole physics. 
%We analytically study the one-loop correction to the black hole thermodynamics in a restricted phase space; higher-loop corrections are beyond the scope of the present work.
%The situation when subleading-order of $G_\tN$ corrections to black hole thermodynamics are considered should be regarded as ``quantum (or loop) corrected black hole thermodynamics".
It is worth emphasizing that the quantum corrections considered in this work should be understood in a restricted but physically motivated sense. We focus on the off-shell black hole geometries in the free energy landscape that are parametrized by the conical singularities and the black hole size that distinguish the small and large black hole branches. Thus, the parameter $r_\hh$ plays the role of a collective variable for the thermodynamic phase structure, and the reduced path integral over this collective sector captures the most important black hole saddles. %In this sense, the resulting corrections are quantum corrections associated with the horizon radius collective mode.

It can be shown that the quantity $\bar{\Phi}$ conjugating to the given charge $Q$ and $\bar{V}$ conjugating to the given pressure $P$, can be derived with respect to the given Gaussian distribution. 
We have
\bea
\bar{\Phi}=\ave{\Phi} &=& \frac{1}{Z}\int e^{-I_\tE} \left(\frac{\partial_Q I_\tE}{\beta} \right)~\left( \partial_{r_\hh} E \right)~\dd r_\hh \nn\\
&\approx & %\frac{1}{\beta}\frac{\partial I_\tE}{\partial Q}+\frac{1}{2 \beta  I_{\tE}''}\cdot \frac{\partial I_{\tE}''}{\partial Q} + \mathcal{O}(G_\tN^1)\nn\\ &=& 
\frac{1}{\beta}\partial_Q \left[ I_\tE+\frac{1}{2} \ln I_{\tE}'' \right]+ \mathcal{O}(G_\tN^1)\,,
\eea
and 
\bea
\bar{V}=\ave{V} &=& \frac{1}{Z}\int e^{-I_\tE} \left(\frac{\partial_P I_\tE}{\beta} \right)~\left( \partial_{r_\hh} E \right)~\dd r_\hh \nn\\
&\approx & % \frac{1}{\beta}\frac{\partial I_\tE}{\partial P}+\frac{1}{2 \beta  I_{\tE}''}\cdot \frac{\partial I_{\tE}''}{\partial P} + \mathcal{O}(G_\tN^1)\nn\\ &=& 
\frac{1}{\beta}\partial_P \left[ I_\tE+\frac{1}{2} \ln I_{\tE}'' \right]+ \mathcal{O}(G_\tN^1)\,.
\eea
The above $\bar{\Phi}$ and $\bar{V}$ exactly match the quantities being derived from the effective action, i.e.
\be
\bar{\Phi}= \frac{1}{\beta} \partial_Q \Gamma_\eff\,,~~~~~~~~\bar{V}= \frac{1}{\beta} \partial_P \Gamma_\eff\,.
\ee
It is interesting to notice that the averaged free energy should be reinterpreted as the potential conjugate to Newton's constant
\bea
\bar{\mu}=\ave{F} &=& \frac{1}{Z}\int e^{-I_\tE} \left(\frac{I_\tE}{\beta} \right)~\left( \partial_{r_\hh} E \right)~\dd r_\hh \nn\\
&\approx & - \frac{G_\tN}{\beta}\partial_{G_\tN} \left[ I_\tE+\frac{1}{2} \ln I_{\tE}'' \right]+ \mathcal{O}(G_\tN^1)\,.
\eea
This means that
\be
\frac{\bar{\mu}}{G_\tN}=- \frac{1}{\beta}\partial_{G_\tN} \Gamma_\eff\,.
\ee
Note that $G_\tN$ is the dimensionless Newton's constant, so $1/G_\tN$ here can be regarded as the center charge $c\propto N^2$ in the dual boundary theory. Our interpretation here should match the black hole thermodynamics with varying center charge \cite{Visser:2021eqk,Cong:2021fnf,Ahmed:2023snm,Ahmed:2023dnh,Zhang:2023uay,Mancilla:2024spp}.

The effective entropy can be derived from the effective action as
\be
S_\eff=-(1-\beta\partial_\beta)\ln Z=-(1-\beta\partial_\beta)\Gamma_\eff=\frac{\pi r^2_\hH}{G_\tN}-(1-\beta\partial_\beta)\left(\frac{1}{2} \ln I_{\tE}'' \right)\,.
\ee
The leading-order contribution is exactly the Bekenstein-Hawking entropy, and we have a logarithmic correction as the subleading-order contribution.
Define the effective free energy $F_\eff=\Gamma_\eff/\beta$, the first law can be written as
\bea
\dd F_\eff=S_\eff ~\dd T_\tE+ \bar{\Phi} \dd Q + \bar{V}\dd P -\bar{\mu}~\dd \ln G_\tN\,.\label{dfeff}
\eea
It is easy to show that $\bar{E}$ can be related to $F_\eff$ through a Legendre transformation $F_\eff = \bar{E}-T_\tE S_\eff$, so we also have
\bea
\dd \bar{E}= T_\tE ~\dd S_\eff + \bar{\Phi} \dd Q + \bar{V}\dd P -\bar{\mu}~\dd \ln G_\tN\,.
\eea

The first law of black hole thermodynamics is the essential relation. All the other thermodynamic laws can be verified without much effort. With well-defined thermodynamic laws, we can say that we have valid loop-corrected black hole thermodynamics.

%It is worth emphasizing that 
%\be
%2F_\eff=2S_\eff ~T_{\tE}+\bar{\Phi} Q  +2\bar{V} P+\frac{r_\hh}{G_\tN}
%\ee

Note that we have gained logarithmic corrections in both the effective entropy and the effective action. 
logarithmic corrections are generally expected for the subleading-order quantum correction, for example \cite{tHooft:1984kcu,Solodukhin:1994yz,Fursaev:1995ef,Mann:1996bi,Solodukhin:2011gn,Domagala:2004jt,Solodukhin:2008dh,Honda:2019cio,Cheng:2023bcv,Cheng:2023cms,Hu:2024ldp,Kapec:2023ruw,Iliesiu:2022onk,Maulik:2024dwq,Modak:2025gvp,Ghosh:2023psh}.
It is worth emphasising that the logarithmic correction here 
differs slightly from the above models. 
Taking the``Brick wall'' model \cite{tHooft:1984kcu} and the ``quantum entropy'' model
\cite{Solodukhin:1994yz,Fursaev:1995ef,Mann:1996bi,Solodukhin:2011gn} as examples, the logarithmic correction captures the quantum correction from the one-loop effect of matter on a fixed background. 
In this paper, the logarithmic correction comes from ``quantum geometries'' that are off-shell.

The Smarr relation should not be satisfied when logarithmic corrections are included. This is because when the homogeneousness is lost, the magics in Euler's scaling argument \cite{Caldarelli:1999xj,Kastor:2009wy,Kubiznak:2016qmn} do not work anymore.
The Smarr relation usually means that the expressions of the quantities are simple enough to be manipulated in any way you wish.

\section{Phase transition in the quantum-corrected thermodynamics}
\label{sec:PT}

In this section, we will use the model studied in the previous section to demonstrate the phase transition behavior for black hole systems with logarithmic corrections in black hole entropy and effective free energy.
Note that the effective free energy $\Gamma_\eff/\beta$ depends on the value of $G_\tN$. 
For extremely small $G_\tN$, the logarithmic correction is negligible. With relatively larger $G_\tN$, the correction becomes more and more visible.

\begin{figure*}[hbt]
\centering
\subfloat[The effective free energy with different $G_\tN$ for $P<P_\text{c}$.\label{varyGN}]{\includegraphics[width=0.45\linewidth]{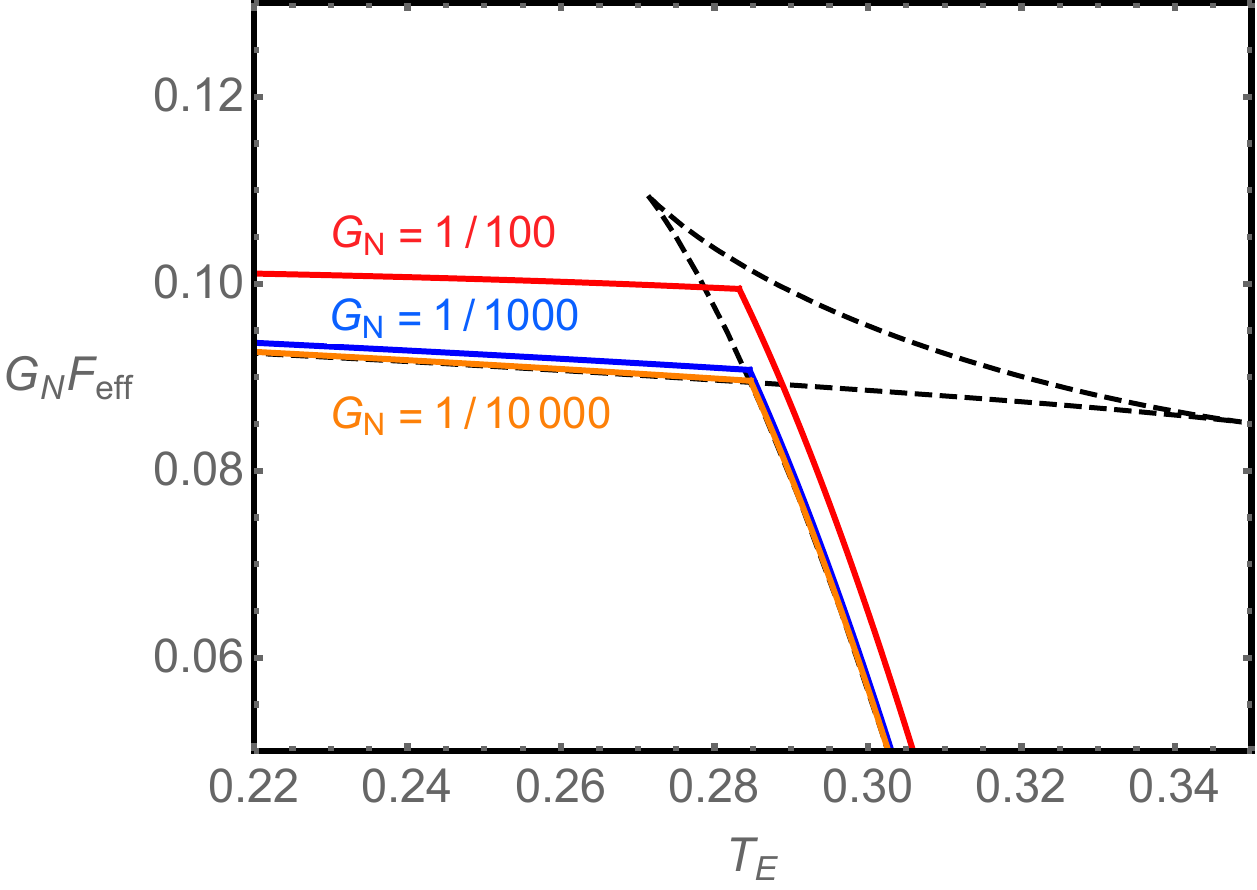}}~~~~\subfloat[The cases with different $G_\tN P$ when $G_\tN=1/1000$.\label{diffP}]{\includegraphics[width=0.47\linewidth]{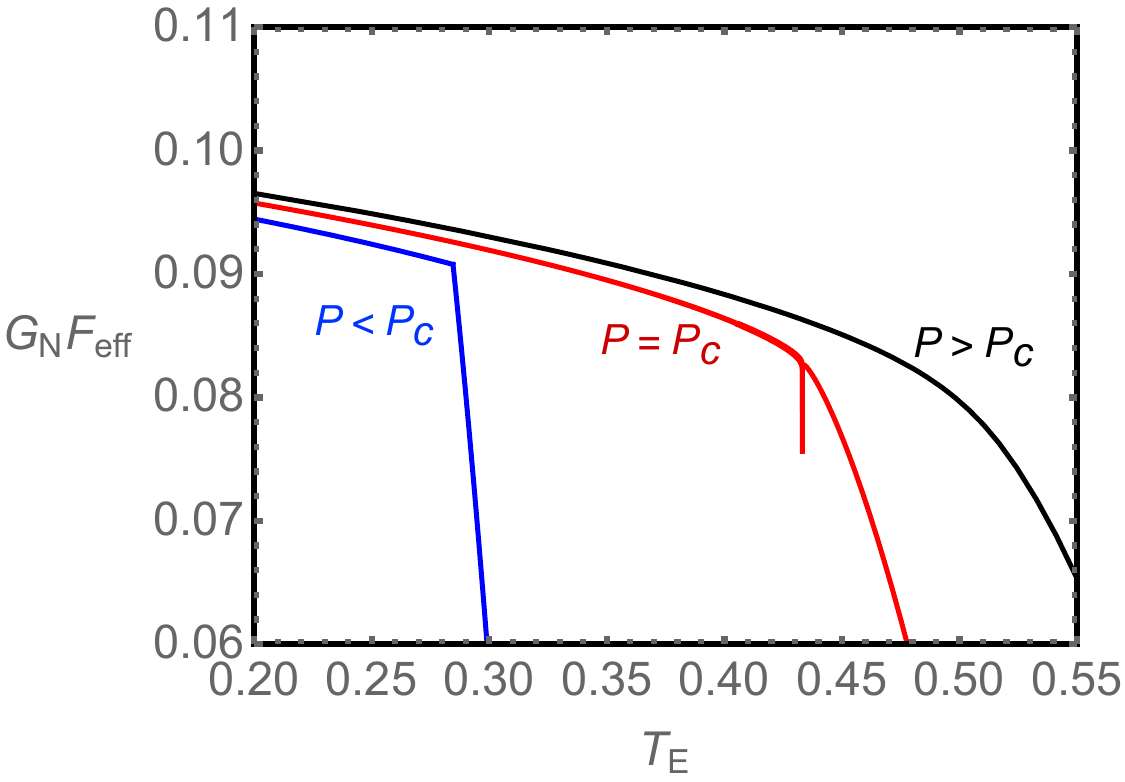}}
    \caption{(a) We have set $Q=0.1$ and $L=1$ in the figure.
     The orange, blue, and red curves respectively correspond to the cases with $G_\tN=1/10000,~1/1000$, and $1/100$, while the dashed swallow-tail curve is the original Gibbs free energy.
    (b) The blue curve illustrates the first-order phase transition with $G_\tN P=0.30<G_\tN P_\text{c}$. The red curve shows the second-order phase transition with $G_\tN P=G_\tN P_\text{c}$. For $G_\tN P=0.48>G_\tN P_\text{c}$, there is no phase transition as shown by the black curve.
    }
    \label{Feff}
\end{figure*}

As shown in Fig. \ref{Feff}, setting $Q=0.1$ and $L=1$, the original Gibbs free energy exhibits swallow-tail behaviour, implying a first-order phase transition for $P<P_\text{c}$. This is demonstrated by the black dashed curve.
When the logarithmic corrections from the previous section are considered, we can see deviations from the dashed curve. 
For example, when $G_\tN=1/1000$, the effective free energy with different $G_\tN P$ is illustrated in Fig. \ref{diffP}.
As demonstrated by the figures, for $P<P_\text{c}$, there is a first-order phase transition, while for $P=P_\text{c}$, a second-order phase transition happens as shown by the red curve in Fig. \ref{diffP}.
The value of the order parameter $r_\hh$ is continuous along the red curve, but the first-order derivative is discontinuous. For larger pressure $P$, everything is continuous and there is no phase transition.
Let us focus on the regime of small $G_\tN P$ where first-order phase transitions occur, as shown by the orange, blue, and red curves in Fig. \ref{varyGN}. 
We only show the phases with lower free energy in the diagram. However, one should keep in mind that, similar to the uncorrected case, the phase transitions are still first-order transitions.
When $G_\tN=1/100$, the deviation is pretty obvious as can be seen from the red curve. However, in the semi-classical limit, the orange curve with $G_\tN=1/10000$ is very close to the original Gibbs free energy.
This implies that our quantum-corrected thermodynamics should have a valid semi-classical limit.

\begin{figure*}[hbt]
\centering
 \subfloat[Phase diagram in the semi-classical limit\label{semi-1}]{\includegraphics[width=0.45\linewidth]{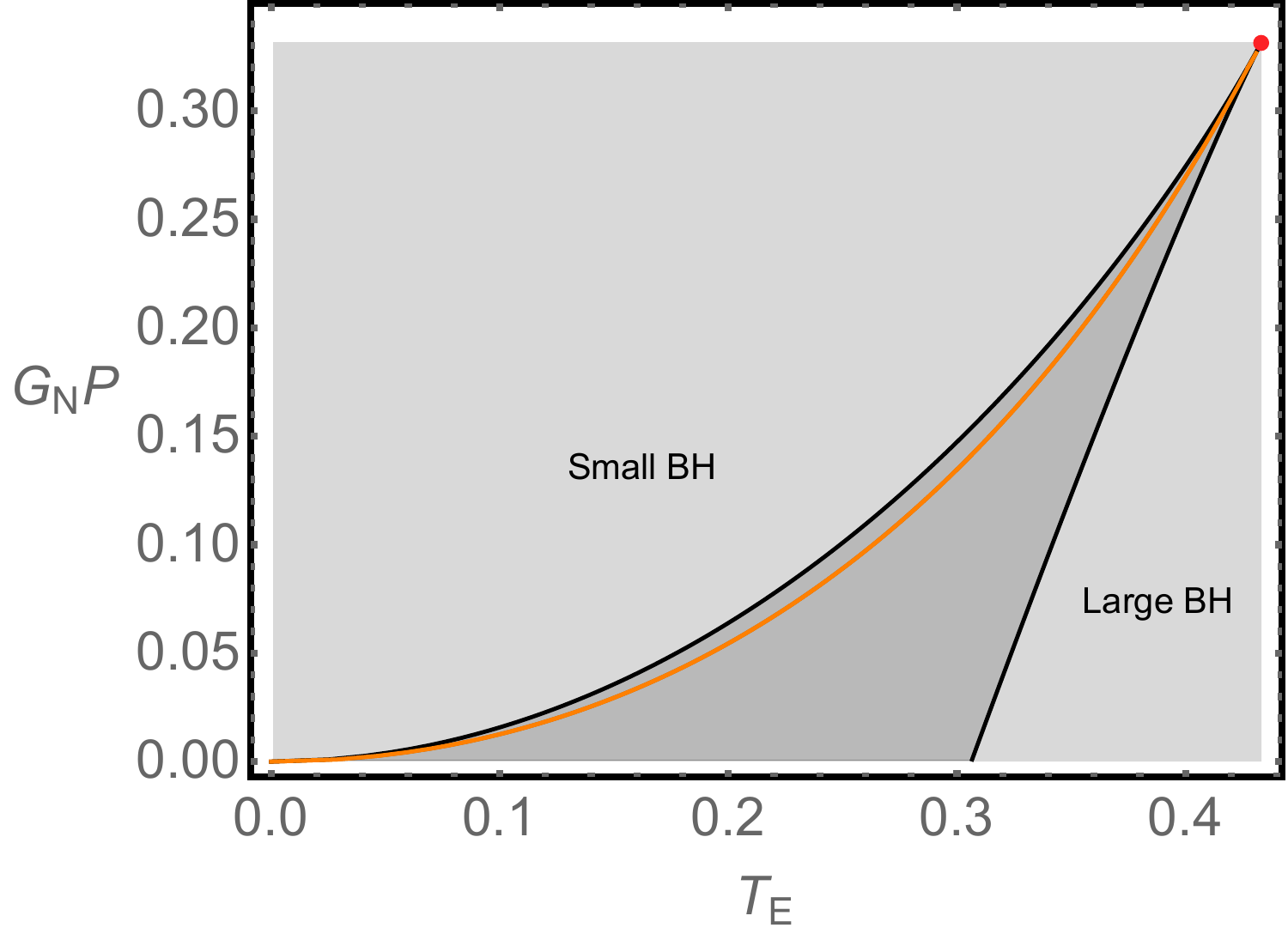 }}
 ~~
  \subfloat[Different phases with $L=1$\label{semi-2}]{\includegraphics[width=0.44\linewidth]{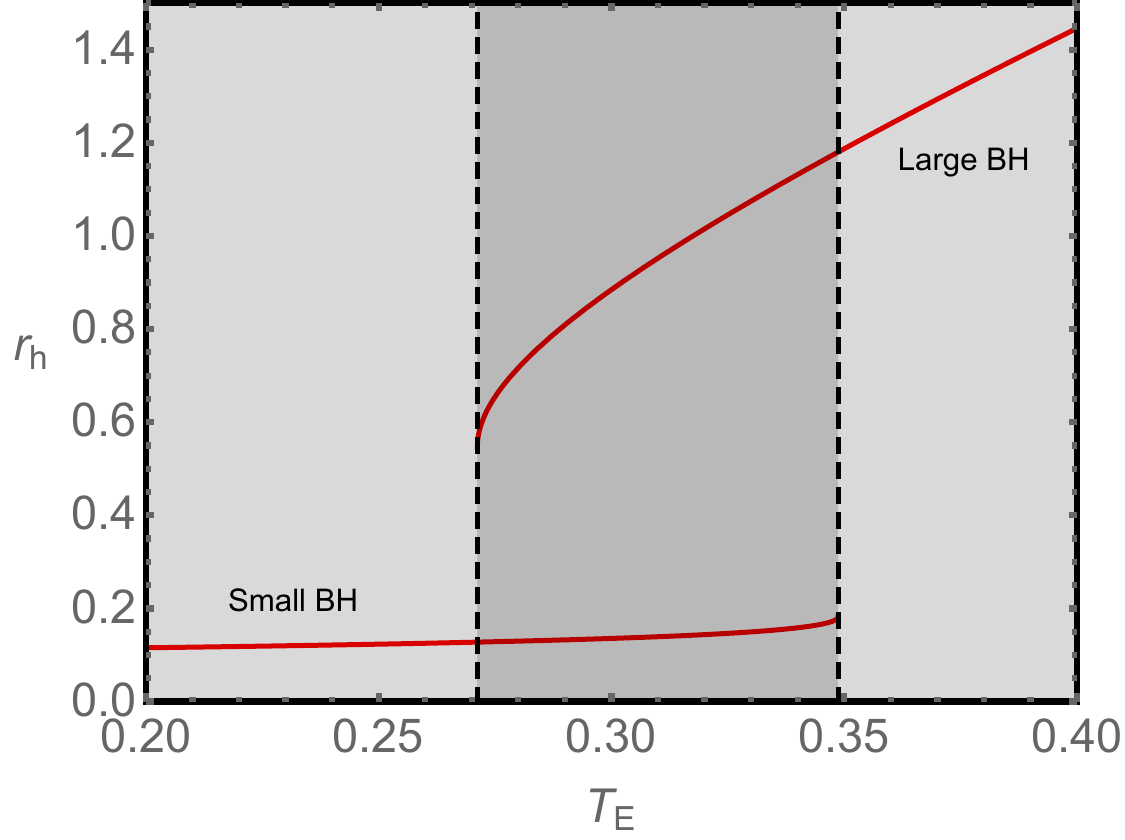 }}
    \caption{Phase transition in the semi-classical limit with $Q=0.1$. (a) illustrates the phase diagram, where the orange line illustrates the coexistence line, and the critical point is denoted by the red dot. The edges between different regions are the spinodal lines. (b) represents different phases with $L=1$. In the two diagrams, the regions with the same grayscale correspond to the same phase.}\label{semi-cl}
\end{figure*}

\subsection{The semi-classical limit}

It is very enlightening to look more closely at the semiclassical limit, where we should be able to recover the traditional black hole thermodynamics completely.

The phase diagram of the black hole phase transition is shown in Fig. \ref{semi-cl}.
As can be seen from Fig. \ref{semi-1}, the coexistence line is illustrated by the orange color. And the critical point corresponding to the second-order phase transition is represented by the red dot.
We have a small black hole phase above the coexistence line and a large black hole phase below the line.
As an example, we take $L=1$ and show the corresponding phases in Fig. \ref{semi-2}.
The shaded regions with the same greyscale correspond to the same phase in the two diagrams.
The region shaded with the darker color demonstrates the situation when there are non-zero possibilities of having small or large black holes, as previously discussed in section \ref{review}.
The phase with negative specific heat is not shown in Fig. \ref{semi-2}. This is because the effective free energy takes complex values for negative specific heat. In such a sense, the unphysical states are eliminated when the corrected effective free energy is considered.

All the physics discussed here should agree with what has been illustrated in section \ref{IIsemi}. For extremely small $G_\tN$ ($G_\tN<1/10000$ in the semi-classical limit), the probability distribution can be well approximated by the Dirac delta function, so the dominant contribution arises from the Euclidean geometry without conical singularity. This is exactly the situation being handled in traditional black hole thermodynamics.

It is worth noticing that for the original black hole thermodynamics, it is not necessary to plot the region shaded with the darker color. All the regions above or below the coexistence line are regarded as the same phase. 
In the logarithmically corrected situation studied in this paper, the darker region can be regarded as a phase with the superposition of different black hole configurations.
Especially, the edges between the regions with different greyscale values are the spinodal lines, which are illustrated by the black curves in Fig. \ref{semi-1}.
The effective free energy is divergent along the black curves, which is not the case for the traditional black hole thermodynamics.
The divergence comes from the logarithm of the specific heat, so this behavior should be common for the effective free energy with logarithmic corrections.

\begin{figure}[hbt]
\centering
\includegraphics[width=9.5cm]{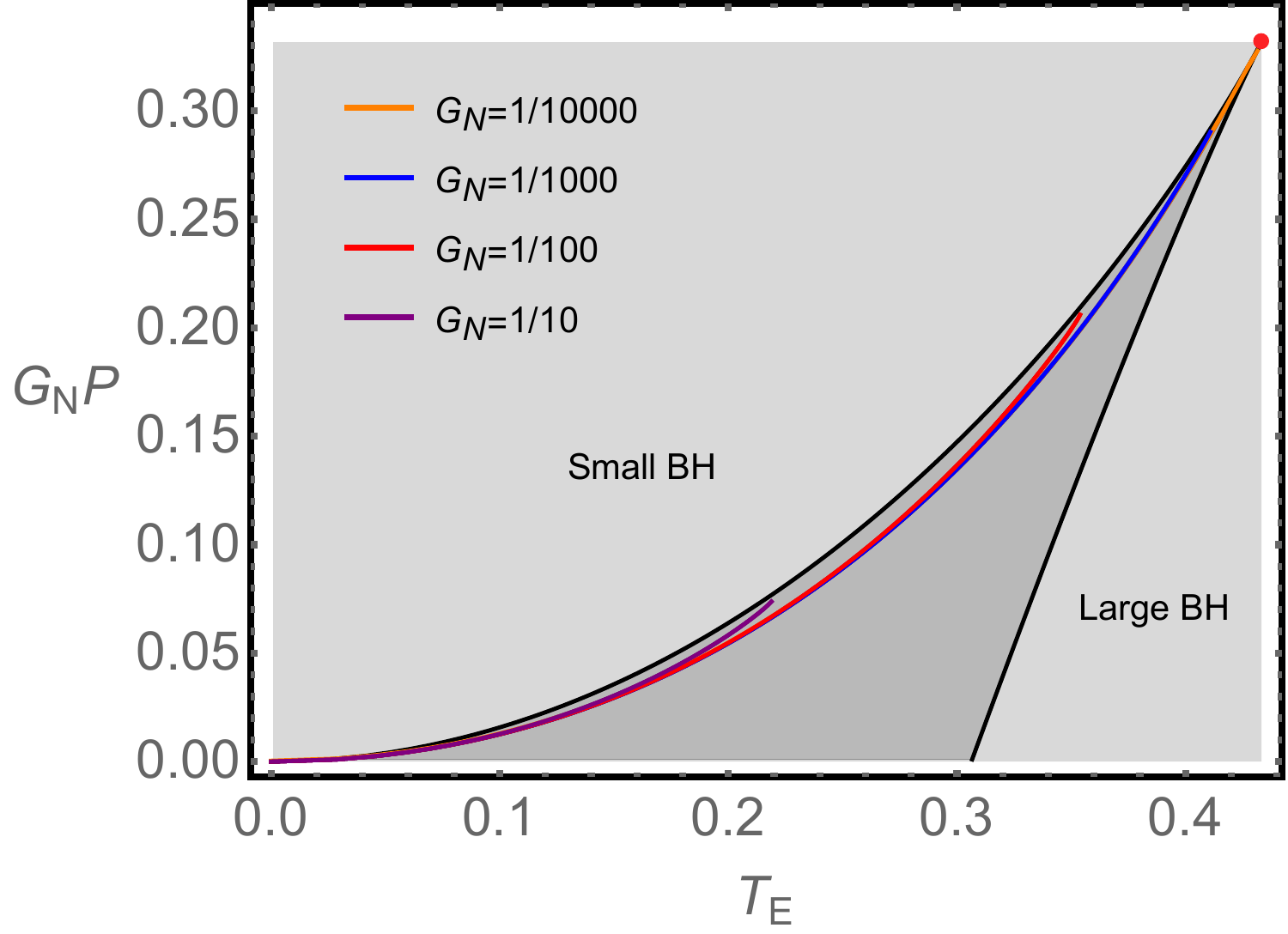}
    \caption{Phase diagrams with different $G_\tN$. We have set $Q=0.1$, and different values of $G_\tN$ are represented by purple, red, blue, and orange colors. The spinodal lines are represented by black curves.}
    \label{phase-diagram}
\end{figure}

\begin{figure*}[hbtp]
\centering
\subfloat[Phase diagram with $G_\tN=1/500$
 \label{G-500}]{\includegraphics[width=0.45\linewidth]{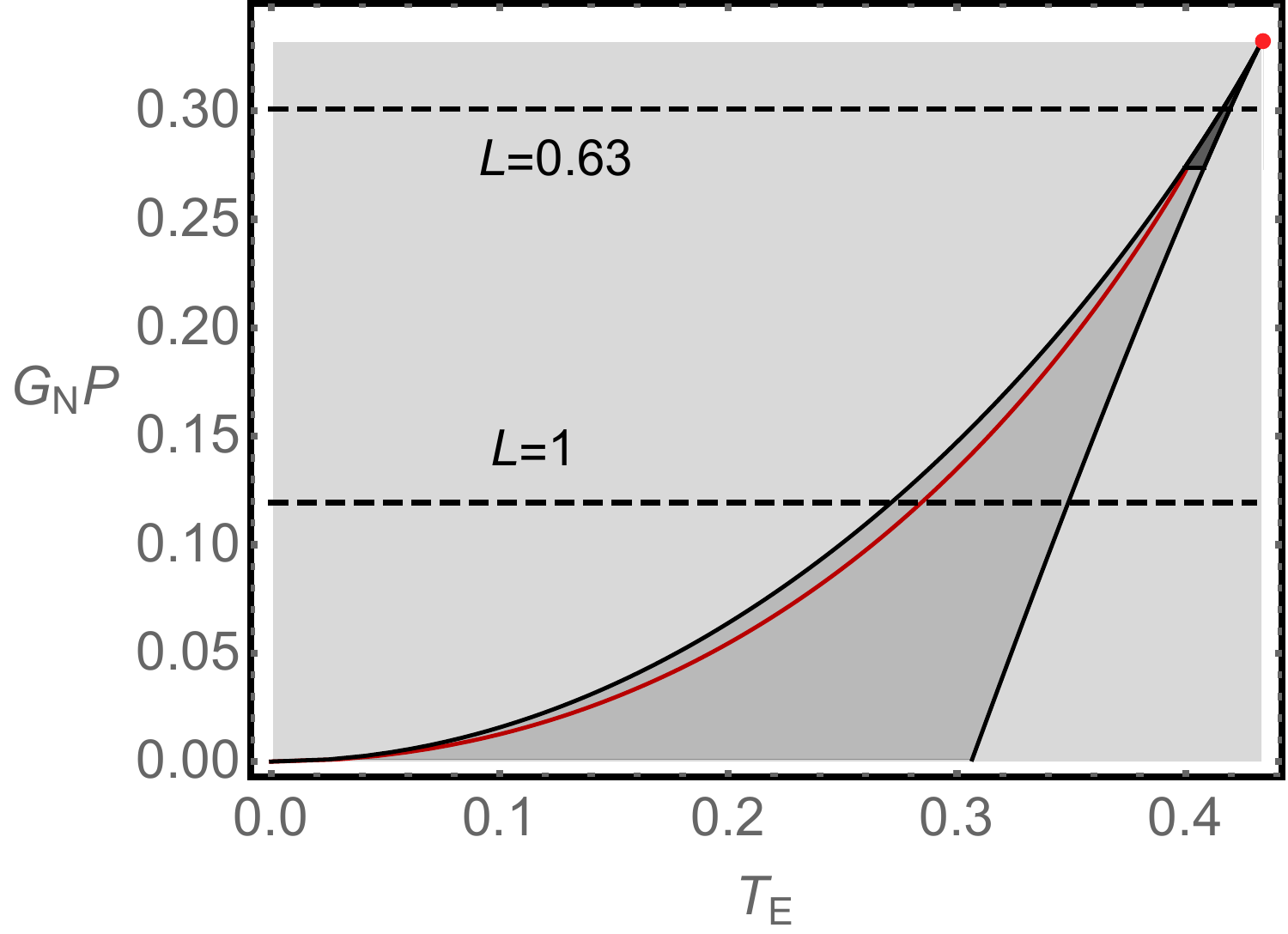 }}
 ~~~
 \subfloat[Phase diagram with $G_\tN=1/100$
 \label{G-100}]{\includegraphics[width=0.45\linewidth]{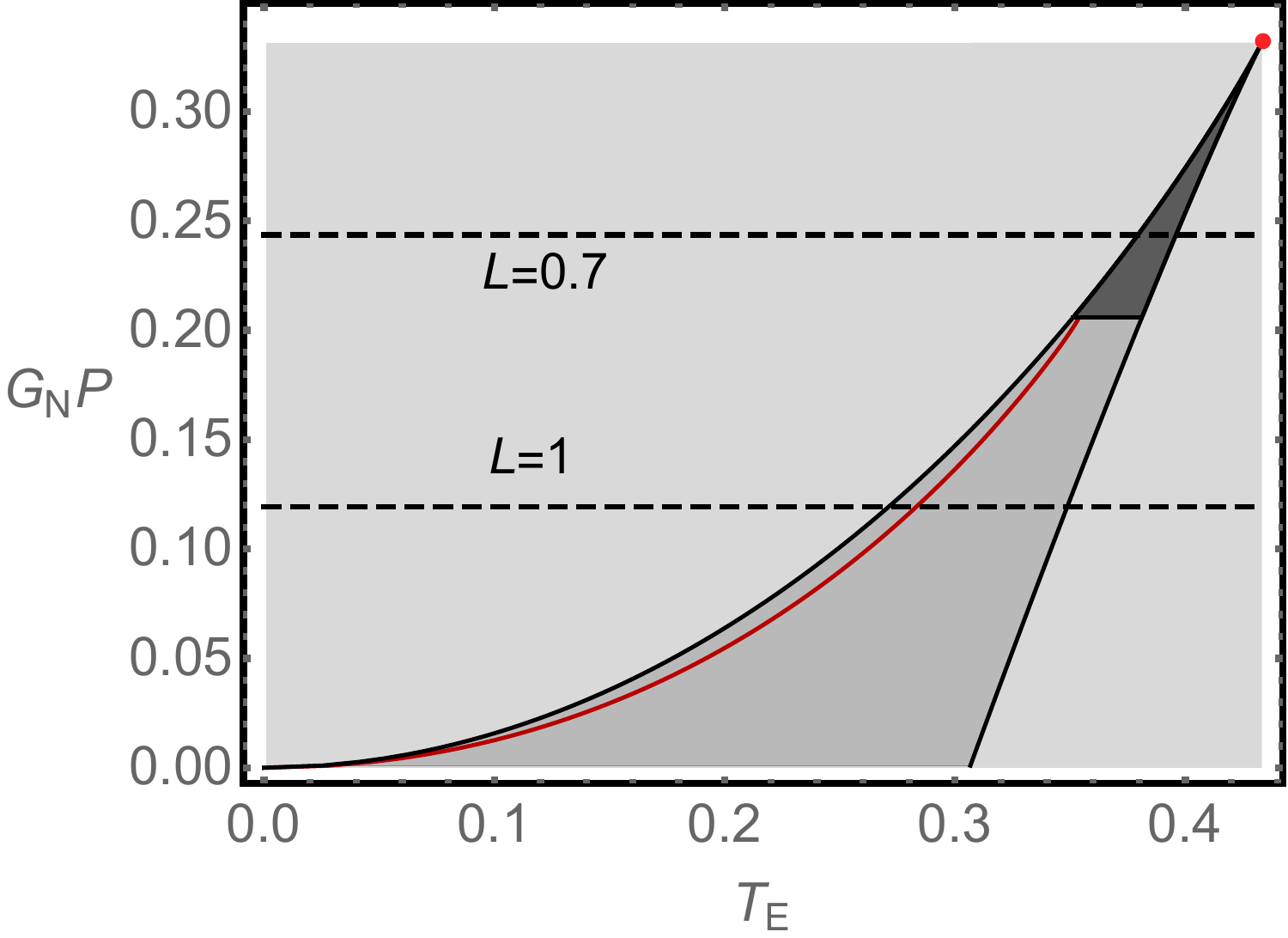 }}
 \\
 \subfloat[For $G_\tN=1/500$, the effective free energy with $L=1$
 \label{L-1}]{\includegraphics[width=0.47\linewidth]{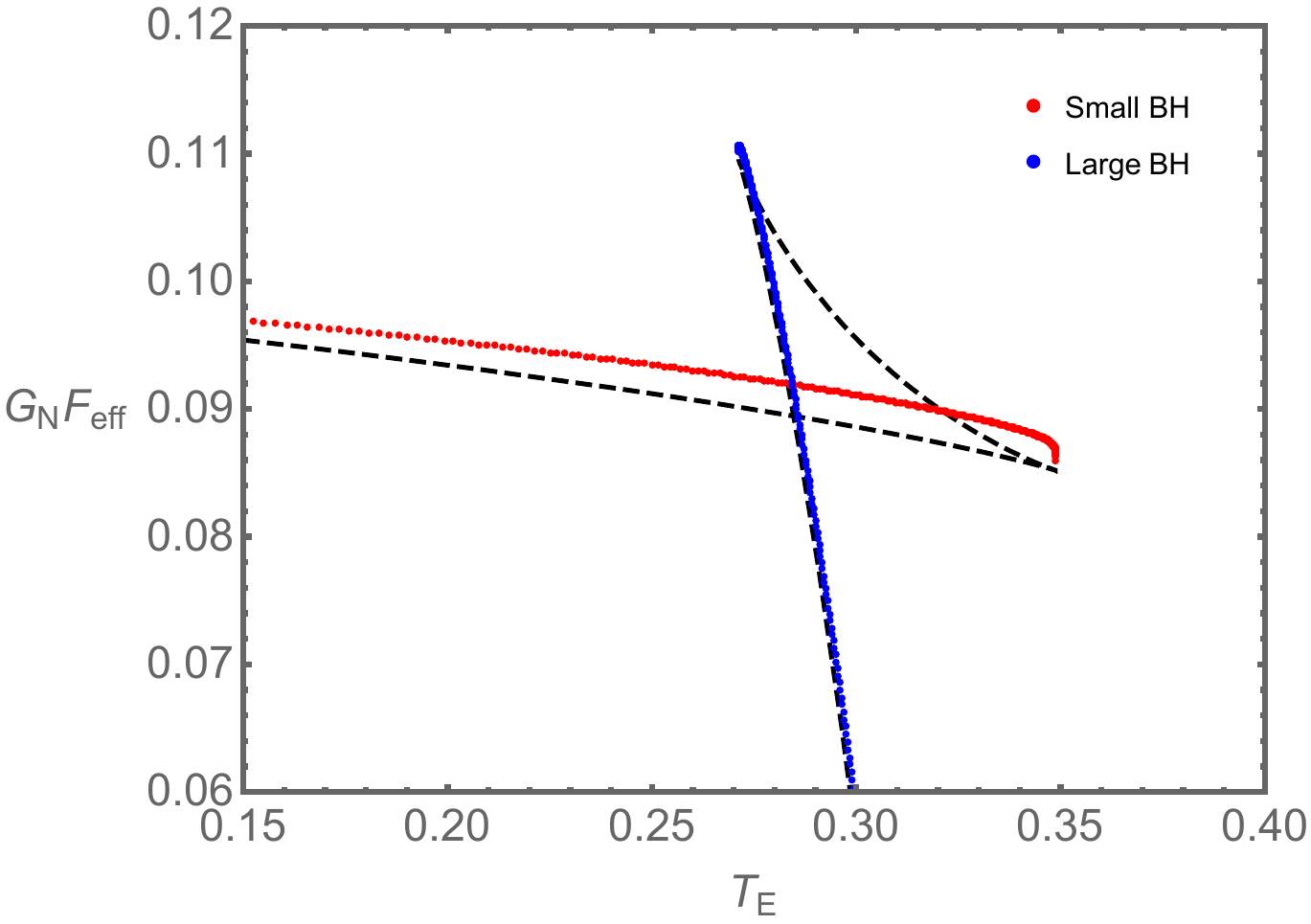 }}~~
 \subfloat[For $G_\tN=1/100$, the effective free energy with $L=1$
\label{L-100-1}]{\includegraphics[width=0.47\linewidth]{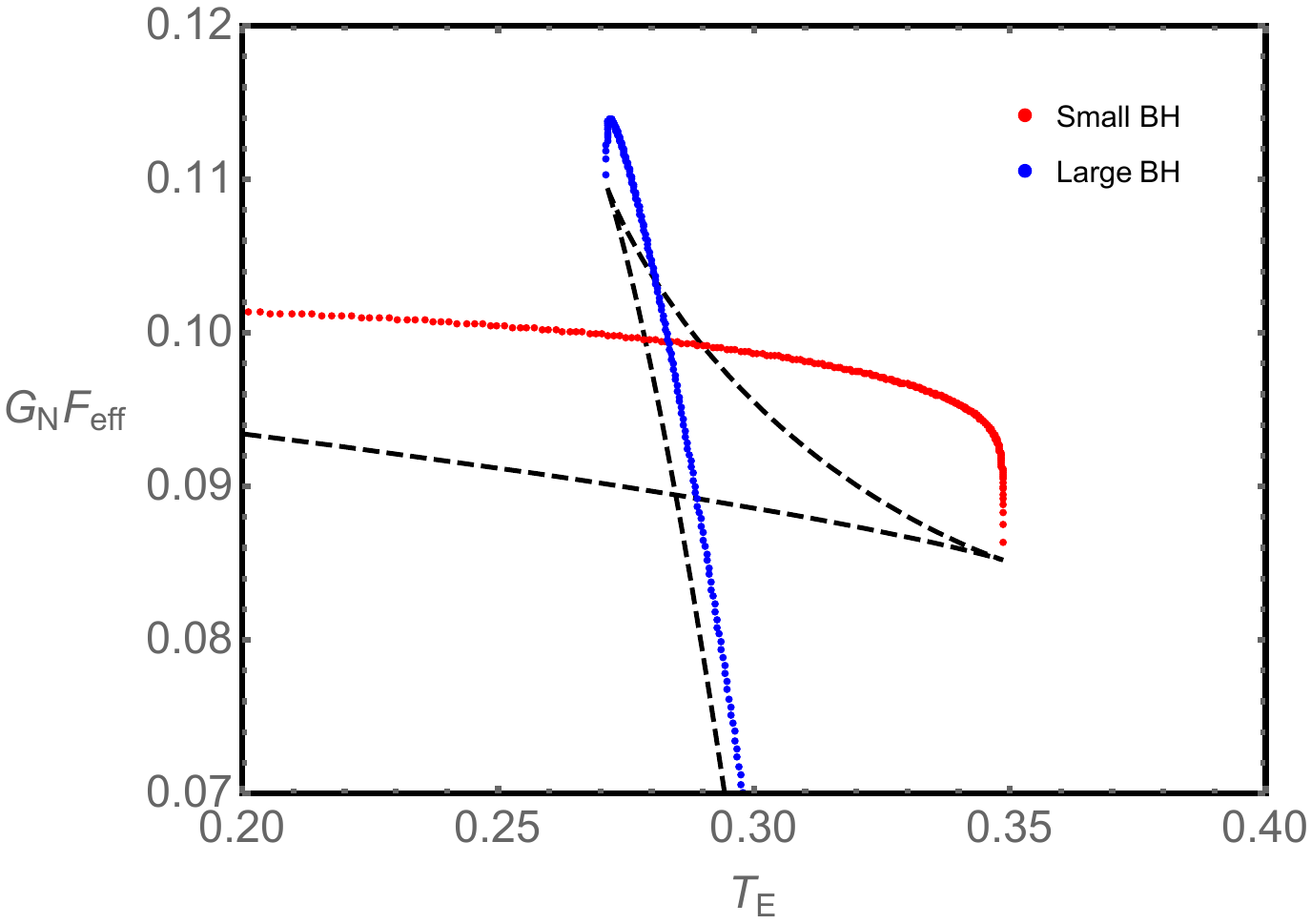}}\\
\subfloat[For $G_\tN=1/500$, the effective free energy with $L=0.63$
\label{L-0.63}]{\includegraphics[width=0.47\linewidth]{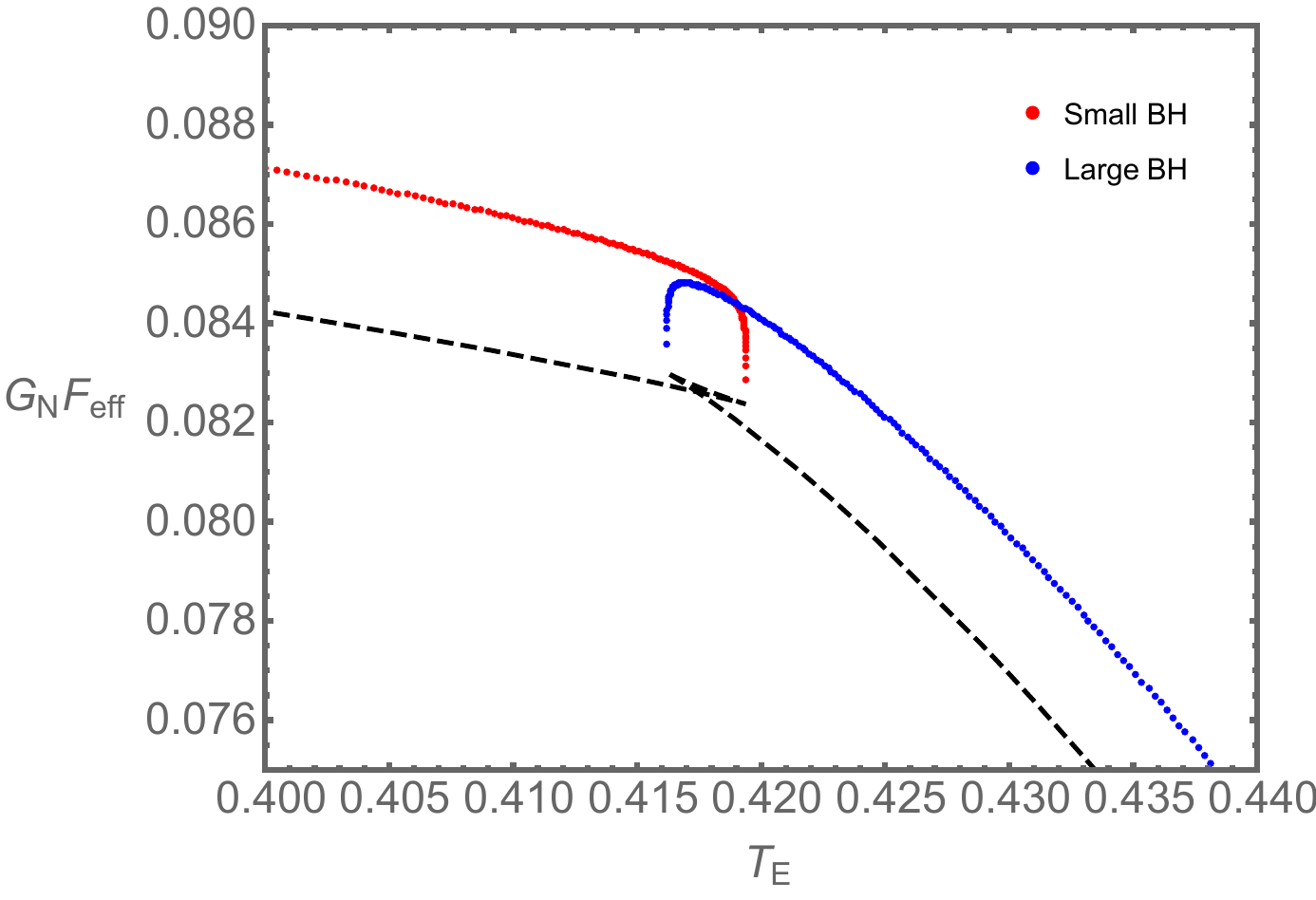}}~~
 \subfloat[For $G_\tN=1/100$, the effective free energy with $L=0.7$
\label{L-100-0.7}]{\includegraphics[width=0.47\linewidth]{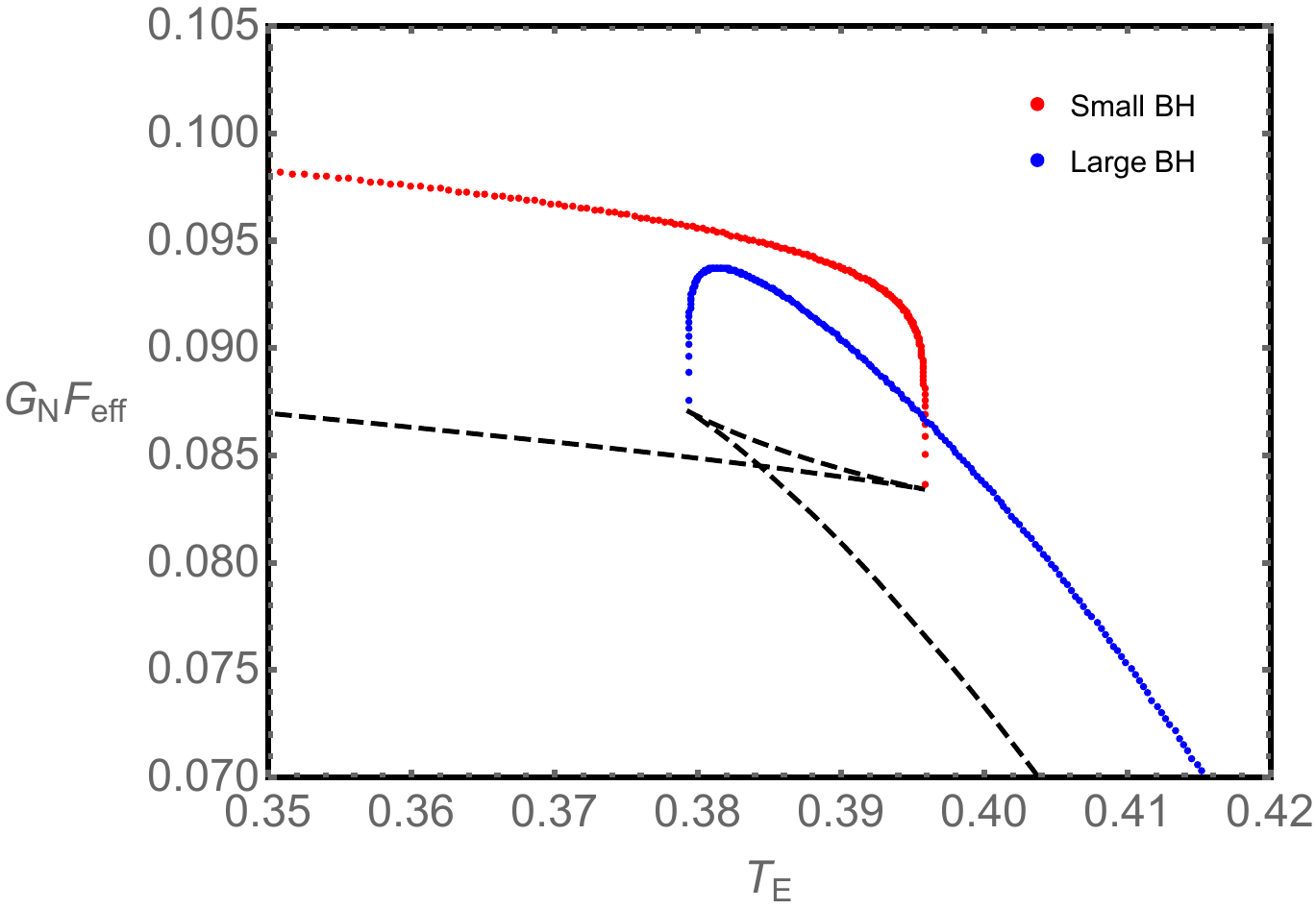}}
    \caption{The two uppermost figures are phase diagrams with different $G_\tN$ ($G_\tN=1/500$ for diagram (a) and $G_\tN=1/100$ for diagram (b)). The four diagrams below show the corresponding effective free energy with different $L$. (c) and (e) are cases with $G_\tN=1/500$ where we have $L=1$ and $L=0.63$ respectively; When $G_\tN=1/100$, the diagrams are shown in (d) with $L=1$ and (f) with  $L=0.7$.}
    \label{500and100}
\end{figure*}

\subsection{Phase tranistion with finite $G_\tN$}

For finite $G_\tN$, the phase diagrams describing the transition between the small and large black holes can be obtained without much effort. 
The phase diagrams are shown in Fig. \ref{phase-diagram}.
For different $G_\tN$s, the critical points where the second-order phase transition happens are always near $(T_\tE \approx 0.433,~ G_\tN P \approx 0.332)$.
For pressure larger than the critical value, there is no phase transition.
The situation is complicated for the pressure smaller than the critical value, as shown in the diagram.
The coexistence lines do not coincide for different values of $G_\tN$.
As shown in Fig. \ref{phase-diagram}, for $G_\tN P=0.05$, the coexistence lines are more partial to the small black hole phase for the larger $G_\tN$.
This means that the phase transition happens at a lower temperature for the larger $G_\tN$.
For finite $G_\tN$, such as $G_\tN=1/100$, there is a gap between the largest value of the coexistence line and the critical point. This is a common phenomenon for the purple, red, and green lines in Fig. \ref{phase-diagram}. 
The reason is that the first-order phase transition can only happen below a certain pressure. Let us denote this pressure as $P_0$.
When the pressure is higher than $P_0$ but lower than $P_\text{c}$, there exist zero-order phase transitions shown in Fig. \ref{L-0.63} and \ref{L-100-0.7}.
As shown in Fig. \ref{phase-diagram}, the value of $P_0$ decreases for larger $G_\tN$.

More detailed phase diagrams are shown in Fig. \ref{500and100}, where we have set $G_\tN=1/500$ for diagram \ref{G-500} and $G_\tN=1/100$ for diagram \ref{G-100}.
In the figure, we have chosen two different values of $L$ that correspond to the situations when we have the first-order phase transition and the zero-order phase transition, and the pictures just right below the phase diagrams show the corresponding effective free energy in Fig. \ref{500and100}.
In those figures, the small black hole phases are represented by the red curves, and the large black holes are shown by the blue curves.
For diagrams \ref{G-500} and \ref{G-100}, the red points represent the critical points where we have second-order phase transitions.
The regions filled with the darkest color correspond to the situations when we have zero-order phase transitions, as demonstrated in Fig. \ref{L-0.63} and \ref{L-100-0.7}.
There are first-order phase transitions in the regions that contain the coexistence lines. 
For $L=1$, the effective free energies demonstrating the first-order transitions are shown in Fig. \ref{L-1} and \ref{L-100-1}. 

Note that the effective free energy itself is discontinuous for the zero-order phase transition.
The region containing the zero-order phase transition becomes larger for larger $G_\tN$. This is because, for larger $G_\tN$, the deviation from the semi-classical limit is more obvious, and the free energies of the two phases are more easily disconnected with large $G_\tN$.
The zero-order phase transition is not commonly described in traditional black hole thermodynamics. %However, we believe this phenomenon should be a universal behaviour when logarithmic corrections are included.
However, our study suggests that zero-order phase transitions may be a generic feature of black hole thermodynamics when logarithmic corrections from off-shell geometries are taken into account.
In the semi-classical realm with very small $G_\tN$, even when the logarithmic corrections are considered, the zero-order phase transition region is too small to be noticed.

%%%%%%%%%%%%%%%%%%%%%%%%%%%%%%%%%%%%%%%%%%%%%%%%%%%%%%%%%%%%%%%%%%%%%%%%%%%%%%%%%%%%%%%%%%%%%%%%%%%%

\section{Conclusion and discussion}
\label{con}

%%%%%%%%%%%%%%%%%%%%%%%%%%%%%%%%%%%%%%%%%%%%%%%%%%%%%%%%%%%%%%%%%%%%%%%%%%%%%%%%%%%%%%%%%%%%%%%%%%%%
In this paper, we have studied the quantum-corrected effective thermodynamics of RN-AdS black holes due to off-shell geometries in a reduced gravitational path integral.
%In this paper, we have studied the quantum-corrected black hole thermodynamics due to the off-shell geometries in the gravitational path integral.
The density matrix and probability distribution of the geometries, both on-shell and off-shell, can be obtained by the Euclidean path integral method.
The probability distribution of the off-shell geometries can be non-zero for finite $G_\tN$, as shown in Fig. \ref{Pdistribution}.
Using the density matrix and the probability distribution, all the physical quantities should be defined as ensemble-averaged ones.
Moreover, in the semi-classical limit $G_\tN \to 0$, the ensemble-averaged quantities recover the quantities we usually see in black hole thermodynamics.

Based on the ensemble-averaged theory, we can obtain the corrected quantities by working out the corresponding integral. 
For relatively small $G_\tN$, expanding the integral around the on-shell configuration is a good approximation. 
This gives us the quantum-corrected quantities.
It can be easily checked that the effective quantities satisfy the four thermodynamic laws; thus, we have a well-defined thermodynamics.
It is worth emphasizing that the effective action consistent with all the corrected quantities is associated with a logarithmic correction.
Correspondingly, there is a logarithm term in the effective entropy and the effective free energy.
The logarithm term originates from the off-shell geometries in the gravitational path integral, rather than quantum matter.

The phase transition behavior is thoroughly analyzed in the corrected thermodynamics.
The phase diagrams are obtained for different $G_\tN$ values characterizing the amount of off-shell geometries included.
In the semi-classical limit, all the phase transition behavior of RN-AdS black holes can be recovered.
However, with larger values of $G_\tN$, the influence of off-shell geometries on the coexistence curve is pretty obvious as shown in Fig. \ref{phase-diagram}. 
The region where we have first-order phase transitions becomes smaller for larger $G_\tN$, and the transitions tend to happen at lower temperatures.
There is an extra region where we have zero-order phase transitions for the RN-AdS black holes.
The critical point where the second-order phase transition happens is not modified.
And the region between the first-order and second-order transition, as shown by the darkest color in Fig. \ref{G-500} and \ref{G-100}, is where the zero-order phase transition happens.
As expected, the deviation from the transitional black hole thermodynamics and the zero-order phase transitions region increases for larger $G_\tN$.
The zero-order phase transitions are explicitly demonstrated through the behavior of the effective free energy in Fig. \ref{500and100}.
Zero-order transition is rare in traditional black hole thermodynamics, but we believe this phenomenon should be common when logarithmic corrections are included, although always been ignored in the literature.

%We are trying to understand the key feature of the quantum corrections, especially their influence on the phase diagrams in this paper.
%The effects of off-shell geometries with conical singularity considered here should be regarded as the vital physics, and more importantly, capture the general properties of off-shell geometries in the gravitational path integral.
%Including off-shell geometries can naturally explain the black hole phase transitions related to the switch of different saddles. Moreover, the density of states with different $E$ in the path integral provides a well-defined statistical interpretation.
%The correction in the  effective action has a universal form, and the reason for the universality is that the subleading-order contribution of $G_\text{N}$ when expanding around the saddle is Gaussian. 
%As proved in \cite{Cheng:2024efw}, the coefficient $1/2$ associated with the logarithmic correction is a signature of the Gaussianity. 
%The universality fixes the qualitative behavior and the correction to the phase diagram.

It is worth mentioning that, at first glance, the critical behaviour illustrated by the red curve in Fig. \ref{diffP} is somewhat strange.
The critical behaviour is different than what we commonly see in standard black hole thermodynamics.
Interesting new physics might be hidden in the second-order phase transition, and one may find a microscopic description of the critical behaviour.
We would like to leave this phenomenon for further studies.

It is important to clarify the scope of the effective description developed in this work. The logarithmic contribution obtained here arises from the reduced path integral over off-shell black hole geometries parametrized by the horizon radius $r_\hh$. In this sense, it captures the quantum correction associated with the horizon-radius collective sector, which is directly related to the black hole free energy landscape and the thermodynamic saddle structure. This contribution should be distinguished from the full one-loop functional determinant of quantum gravity, which would involve local metric, matter, and more fluctuations around a given saddle. The present framework therefore provides an effective and tractable description of the off-shell geometric contribution to black hole thermodynamics. 
It is also worth emphasizing that the thermodynamic landscape connecting different black hole saddles captures the saddle structure most relevant to the gravitational path integral in the regime considered here.
Extending the analysis by including additional collective modes would be an important direction for future work.

%%%%%%%%%%%%%%%%%%%%%%%%%%%%%%%%%%%%%%%%%%%%%%%%%%%%%%%%%%%%%%%%%%%%%%%%%%%%%%%%%%%%%%%%%%%%%%%%%%%%

\section*{Acknowledgements}
We thank Jindong Pan and Si-Jiang Yang for the helpful discussions. 
This work is supported by the National Natural Science Foundation of China (Grant No. 12405073) and Tianjin University Self-Innovation Fund Extreme Basic Research Project (Grant No. 2025XJ21-0007).
%and the Natural Science Foundation of Tianjin(Grant No. xxxx).

%%%%%%%%%%%%%%%%%%%%%%%%%%%%%%%%%%%%%%%%%%%%%%%%%%%%%%%%%%%%%%%%%%%%%%%%%%%%%%%%%%%%%%%%%%%%%%%%%%%%

%\newpage
%\appendix

%%%%%%%%%%%%%%%%%%%%%%%%%%%%%%%%%%%%%%%%%%%%%%%%%%%%%%%%%%%%%%%%%%%%%%%%%%%%%%%%%%%%%%%%%%%%%%%%%%%%

\providecommand{\href}[2]{#2}\begingroup\raggedright\endgroup

\end{document}